\documentclass[%
superscriptaddress,
preprint,
nofootinbib,
amsmath,amssymb,
aps,
pra,
]{revtex4-1}
\usepackage[section]{placeins}
\usepackage{graphicx}
\usepackage{dcolumn}
\usepackage{bm}

\usepackage[version=3]{mhchem}
\usepackage{color}
\usepackage{float}
\usepackage{bigstrut}
\usepackage{booktabs}

\begin{document}
	\sloppy

\title{Effect of Electron Count and Chemical Complexity in the Ta-Nb-Hf-Zr-Ti High-Entropy Alloy Superconductor}

\author{Fabian von Rohr}
\affiliation{Department of Chemistry, Princeton University, Princeton, New Jersey 08544, USA}
\author{Micha\l{} J. Winiarski} 
\affiliation{Faculty of Applied Physics and Mathematics, Gdansk University of Technology, Gdansk 80-233, Poland}
\author{Jing Tao}
\affiliation{Condensed Matter Physics Department, Brookhaven National Laboratory, Upton, New York 11973, USA}
\author{Tomasz Klimczuk} 
\affiliation{Faculty of Applied Physics and Mathematics, Gdansk University of Technology, Gdansk 80-233, Poland}
\author{Robert J. Cava}
\affiliation{Department of Chemistry, Princeton University, Princeton, New Jersey 08544, USA}

\begin{abstract}
	High-entropy alloys are made from random mixtures of principal elements on simple lattices, stabilized by a high mixing entropy. The recently discovered BCC Ta-Nb-Hf-Zr-Ti high entropy alloy superconductor appears to display properties of both simple crystalline intermetallics and amorphous materials, e.g. it has a well defined superconducting transition along with an exceptional robustness against disorder. Here we show that the valence-electron count dependence of the superconducting transition temperature in the high entropy alloy falls between those of analogous simple solid solutions and amorphous materials, and test the effect of alloy complexity on the superconductivity. We propose high-entropy alloys as excellent intermediate systems for studying superconductivity as it evolves between crystalline and amorphous materials.
\end{abstract}

	
	
	\maketitle
	
	\section*{Introduction}
\begin{figure}
			\centering
			\includegraphics[width=\textwidth]{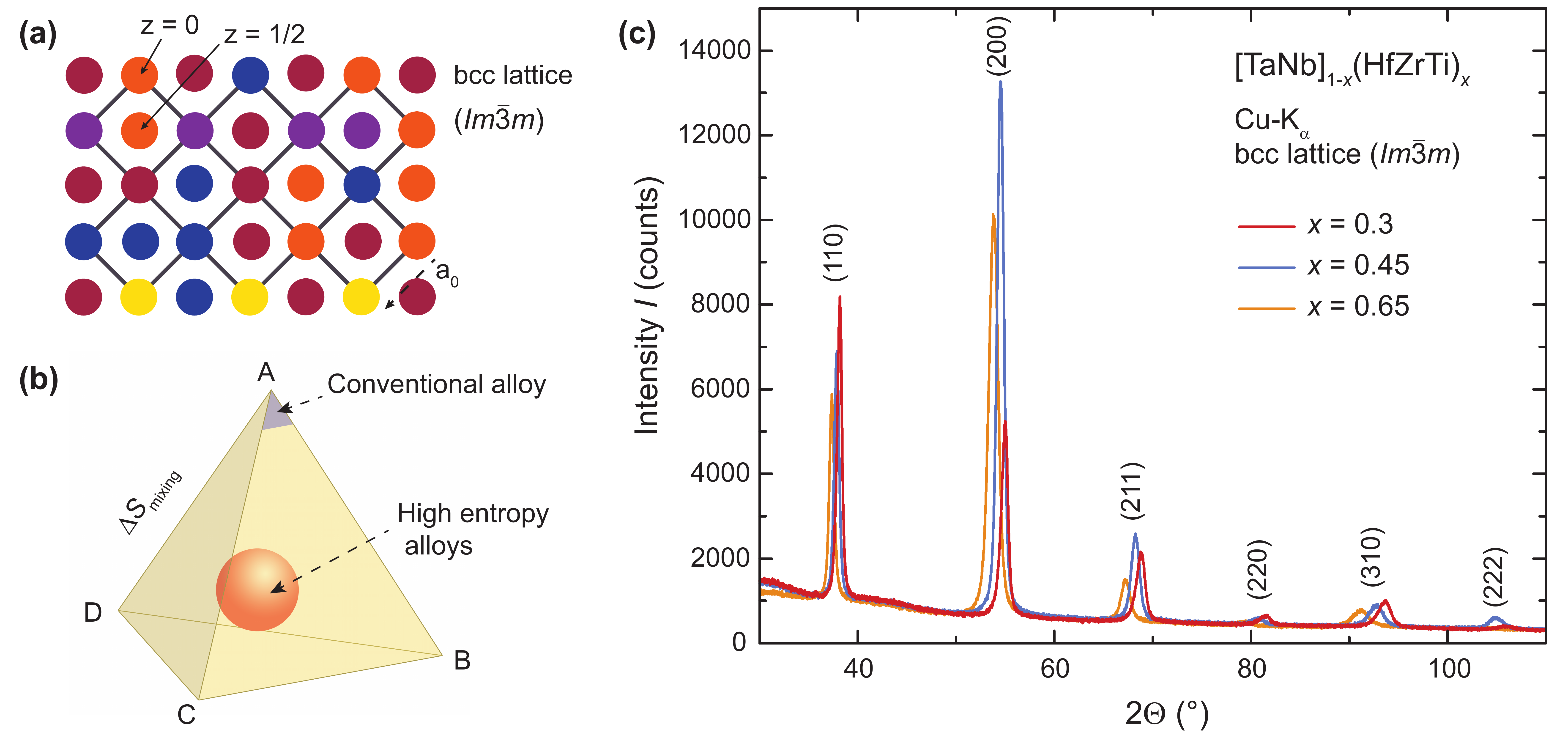}
			\caption{(a) Schematic representation of a BCC lattice with randomly distributed atoms (b) Schematic phase diagram of a multi-component alloy system showing, schematically, conventional and high entropy alloy (HEA) phase regions(c) XRD patterns of the HEAs \ce{[TaNb]_{1-\textit{x}}(ZrHfTi)_{\textit{x}}} for \textit{x} = 0.3, 0.45, 0.65}.
			\label{fig:1}
		\end{figure}

	Alloys are among the most relevant materials for modern technologies. Conventional alloys typically consist of one principal element, such as the iron in steel, plus one or more dopant elements in small proportion (e.g. carbon in the case of steel) that enhance a certain property of interest; the properties are based on the modification of those of the principal element. In sharp contrast, high-entropy alloys (HEA) are comprised of multiple principal elements that are all present in major proportion, with the simple structures observed attributed to the high configurational entropy of the random mixing of the elements on their lattice sites \cite{HEA_Yeh}. Thus, the concept of a "principal element" becomes irrelevant. The elements in HEAs arrange on simple lattices with the atoms stochastically distributed on the crystallographic positions; HEAs are commonly referred to as metallic glasses on an ordered lattice (see figure \ref{fig:1}(a) and (b)). The properties of HEAs arise as a result of the collective interactions of the randomly distributed constituents \cite{HEA_review}. There is no strict definition, but HEAs are typically composed of four or more major elements in similar concentrations. By applying this concept, several new alloys with simple body-centered cubic (BCC), hexagonal-closest packing (HCP), or face-centered cubic (FCC) structures have been realized \cite{HEA_review, Calc_2015}. The HEAs compete for thermodynamic stability with crystalline intermetallic phases with smaller numbers of elemental constituents \cite{HEA_stab}. Therefore, one central concept of designing these alloys is to understand the interplay between mixing entropy $\Delta S_{\rm mixing}$ and phase selection. Considering the large number of metals in the periodic table, the total number of possible HEA compositions is virtually unlimited. 
	\\ \\
	In addition to their structural and chemical diversity, HEAs can display novel, highly-tunable properties such as for example excellent specific strength \cite{HEA_hard,HEA_hard2}, superior mechanical performance at high temperatures \cite{ductile}, and fracture toughness at cryogenic temperatures \cite{cryo,cryo2}, making them promising candidates for new applications. Simple niobium-titanium based binary alloys are nowadays still the most often and widely used materials for superconducting magnets, such as e.g. in NMR and MRI devices \cite{NbTi_sc} or the Large Hardron Collider \cite{LHC}, and thus the discovery of bulk superconductivity with a single well defined phase transition on a highly disordered BCC lattice in the Nb-Ti related Ta-Nb-Hf-Zr-Ti HEA is of considerable interest \cite{HEA_super,HEA_theory}. This multicomponent phase, stabilized by the high mixing entropy, appears to fall between an ordered solid and a glass, and thus allows for study of the chemical composition and structure-property relations of a superconducting material part-way between an ordinary alloy and an amorphous material on a fundamental level. Here, we report the results of our investigations of the influence of electron count and alloy complexity on superconductivity in the Ta-Nb-Hf-Zr-Ti HEA.  We find that the variation in superconducting transition temperature with electron count is intermediate to those displayed by simple alloys and amorphous materials, and that the elemental make-up of the HEA superconductor is critical for determining its properties, in spite of the fact that the materials system is very highly disordered. 
			\begin{figure}
				\centering
				\includegraphics[width=0.4\textwidth]{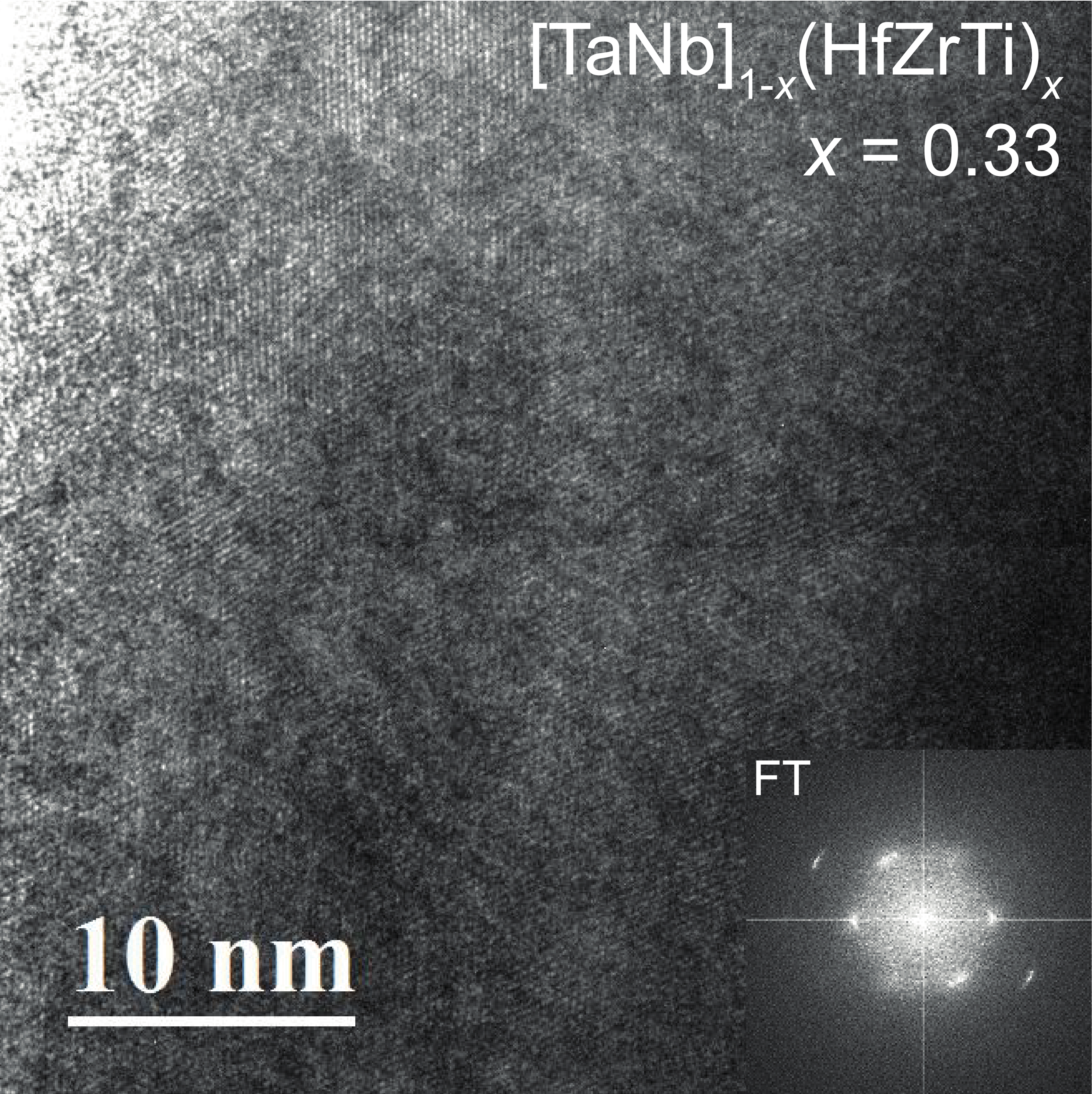}
				\caption{Nano-structure of the HEA \ce{[TaNb]_{1-\textit{x}}(ZrHfTi)_{\textit{x}}} with \textit{x} = 0.33 depicted in a HRTEM image. In the inset, the Fourier-transformation of the observed real space image of the BCC structure, in the [111] zone, is shown.}
				\label{fig:1b}
			\end{figure}  
		\section*{Experimental}
		All samples were prepared from pieces of the pure metals. Stoichiometric amounts of niobium (purity 99.8 \%), tantalum (purity 99.9 \%), zirconium (purity 99.6 \%) hafnium (purity 99.6 \%), and titanium (purity 99.95 \%) pieces were arc melted in high currents  ($T > 2500 \ ^\circ$C) in an argon atmosphere, and rapidly cooled on a water-chilled copper plate. A zirconium sponge was co-heated to purify the reaction atmosphere from remaining oxygen. The samples were melted five times, and turned over each time to ensure optimal mixing of the constituents, the weight loss during melting was found to be insignificant. X-ray diffraction patterns were obtained from mechanically flattened pieces (in liquid nitrogen) of the very hard alloys, measured in a Bragg-Bretano reflection geometry. The patterns were obtained on a Bruker D8 Advance Eco with Cu $K_\alpha$ radiation and a LynxEye-XE detector. The resistivity, magnetization and specific heat were studied using a \textit{Quantum Design} Physical Property Measurement System (PPMS) DynaCool with a 9 T magnet, equipped with a Vibrating Sample Magnetometer Option (VSM). For the resistivity measurements, a standard 4-probe technique was employed with 20 $\mu$m diameter platinum wires attached with silver epoxy. The applied current for these measurements was $I$ = 2 mA. Specific-heat measurements were performed with the Quantum Design heat-capacity option using a relaxation technique. Electron diffraction measurements were performed at Brookhaven National Laboratory on a JEOL ARM200F transmission electron microscope with double-Cs correctors.
	\section*{Results and Discussion}
	\subsection*{Structural Characterization of \ce{[TaNb]_{1-\textit{x}}(ZrHfTi)_{\textit{x}}}}
	The powder x-ray diffraction (XRD) patterns of the high-entropy alloys (HEAs) \ce{[TaNb]_{1-\textit{x}}(ZrHfTi)_{\textit{x}}}\footnote{For better readability of the chemical formula, all elements with a valence electron count (VEC) of 5 are written in squared bracket, while elements with a VEC of 4 are written in round brackets throughout the manuscript.} for \textit{x} = 0.2, 0.25, 0.3, 0.33, 0.35, 0.4, 0.45, 0.5, 0.6, 0.7, 0.8, and 0.84 which were synthesized by arcmelting, can all be indexed with a simple BCC unit cell. All prepared alloys fall within the definition for HEA compositions (see, e.g., reference \onlinecite{HEA_review}), with no constituent element of less than 5 mol-\% and/or more than 40 mol-\%. In figure \ref{fig:1}c, we show three representative XRD patterns of the members $x =$ 0.3, 0.45, and 0.65. The patterns are found to shift only slightly with composition. Therefore a shifting of the cell parameter $a_0$ is observed, but its change between the different HEAs is only minor. All alloys are found to be single phase with broad reflections, which we attribute to both the high degree of disorder present in the HEAs and also the non-ideal diffraction sample preparation (the alloys are too hard to crush by our methods, and so fine particle size powders could not be made for the diffraction experiment). The observed unit cell change results from the large difference of atomic radii of the different constituent atoms. The unit cell parameter for the BCC lattice observed is found to vary from $a_0 \approx 3.33$ \AA \ to $3.43$ \AA \ within the solid solution. The earlier reported cell parameter for variants of this Ta-Nb-Hf-Zr-Ti HEA follow this trend accordingly \cite{HEA_super,Senkov_2011}. Thus the observed physical properties reported below are those of the bulk, since no impurity phases are observed. An earlier reported minor hexagonal phase impurity is not present in the samples of this study \cite{Senkov_2011}. \\ \\
			\begin{figure}
				\centering
				\includegraphics[width=0.5\linewidth]{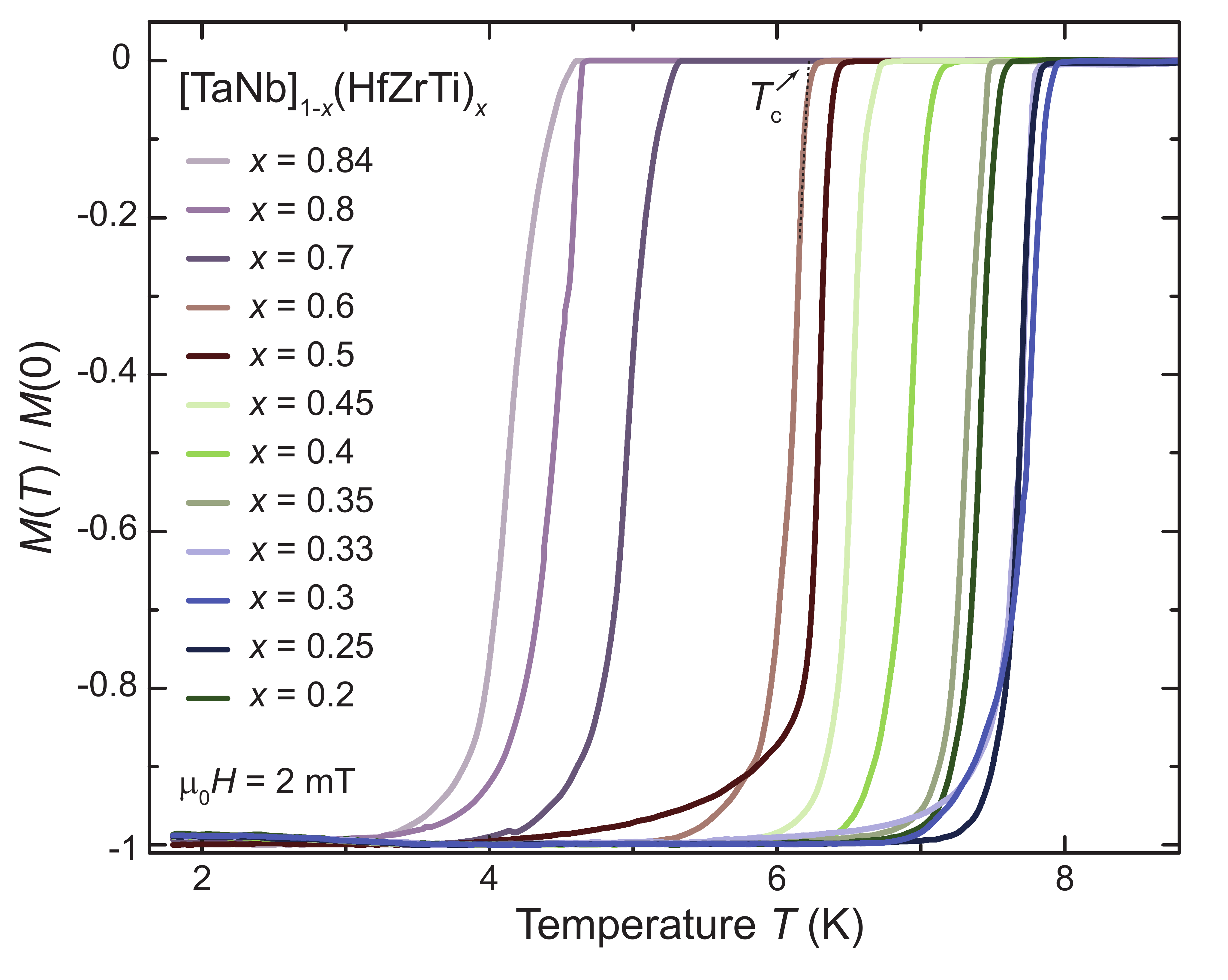}
				\caption{Composition dependence of the superconducting transition in the Ta-Nb-Hf-Zr-Ti high entropy alloy. The ZFC magnetization of the HEAs \ce{[TaNb]_{1-\textit{x}}(ZrHfTi)_{\textit{x}}} with \textit{x} = 0.2, 0.25, 0.3, 0.33, 0.35, 0.4, 0.5, 0.6, 0.7, 0.8, and 0.84 in the vicinity of the superconducting transition, measured in an external magnetic field of $\mu_0 H =$ 2 mT.}
				\label{fig:2a}
			\end{figure}
	In figure \ref{fig:1b}, we show a representative high resolution transmission electron microscope (HRTEM) image of a nearly optimally doped superconducting HEA sample \textit{x} = 0.33. The HRTEM image is taken along the [111] zone axis. This image of the nano-structure of the alloy reveals the arrangement of the atoms on a simple, homogeneous BCC lattice, despite the presence of five constituent atoms with very different atomic radii. Critically, no nanoscale chemical phase separation was observed for any of the materials investigated. In the inset of figure \ref{fig:1b}, we show the Fourier-transform of the observed atom-positions in the real space image. In the Fourier-transform pattern of the HRTEM image, the six reflections close to the center spot represent {110} planes, clearly support the BCC structure of the HEA even at the nanoscale. \\	\\
		\begin{figure}
			\centering
			\includegraphics[width=0.5\linewidth]{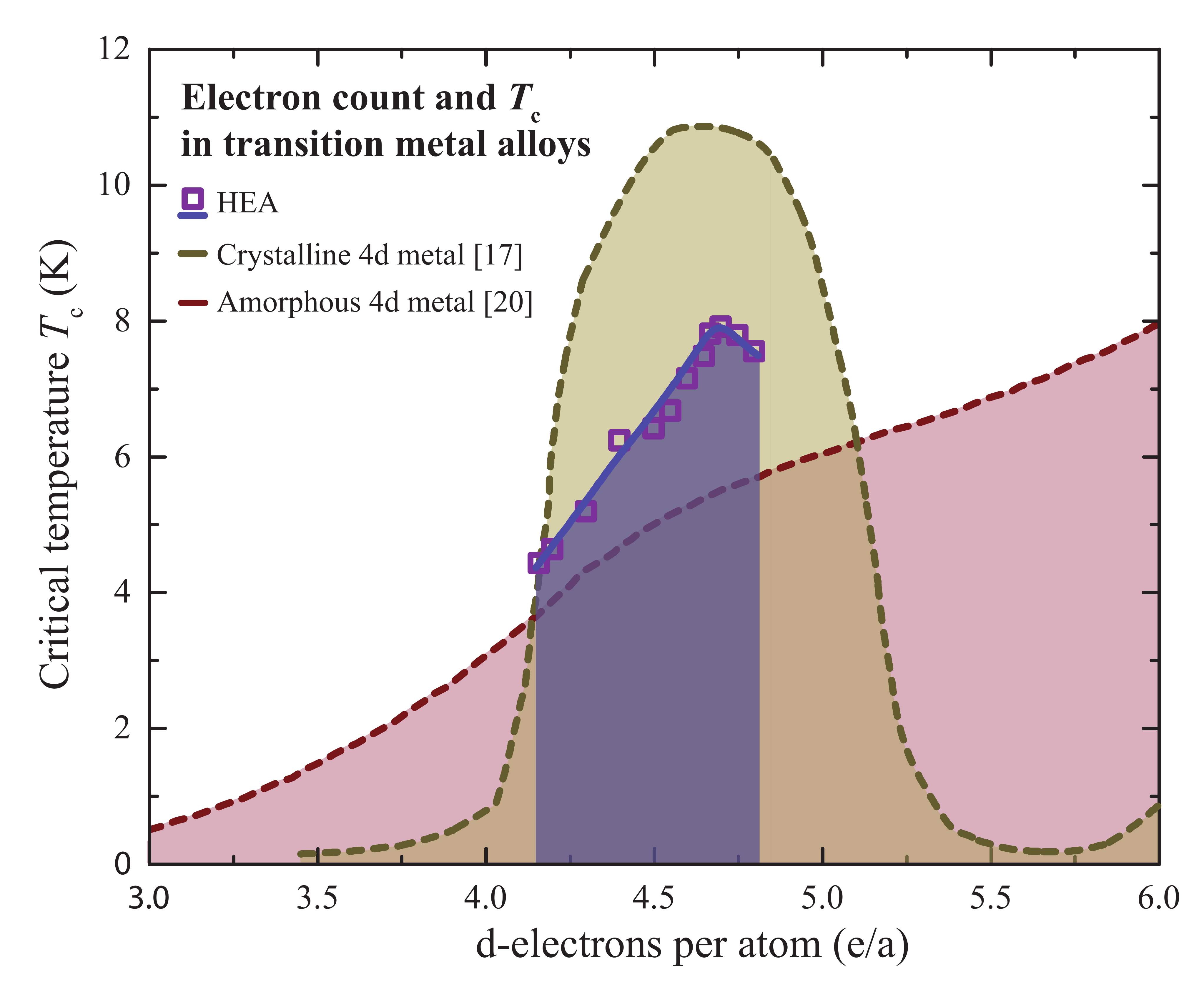}
			\caption{Electron-count dependent superconducting transition temperatures in the high entropy alloy compared to those in analogous simple solid solution and amorphous phases. Phase diagram of \ce{[TaNb]_{1-\textit{x}}(ZrHfTi)_{\textit{x}}} (purple squares are the $T_{\rm c}$ and the blue line is a trend line) in comparison with transition metals and their alloys in the crystalline form \cite{Matthias} (yellow dashed line) and as amorphous vapor-deposited films \cite{amorph1} (red dashed line). The superconducting transition temperatures $T_{\rm c}$ are plotted as function of the electron/atom ratio.}
			\label{fig:2b}
		\end{figure}
	The elemental metals in this pentinary superconducting HEA, when taken by themselves, order on either HCP and BCC lattices: while hafnium, zirconium, and titanium crystallize on a HCP lattice, niobium and tantalum crystallize on a BCC lattice at room temperature. For conventional alloys between metals with a valence electron count (VEC) of 5 (niobium or tantalum) and with a VEC of 4 (titanium, zirconium, or hafnium) a structural transition from a HCP to BCC lattice is observed \cite{martensitic} with decreasing electron count. Due to their electron count, the HEAs prepared here with $x$ = 0.8 and 0.84 would be expected to order on a HCP lattice. This polymorphic transition is, however, not observed in the HEA. The high entropy of the system therefore stabilizes the structure of this HEA preferentially on a BCC lattice (see also, for example references \onlinecite{martensitic,Book_alloys}).
	\subsection*{Electron-Count Dependence of the Superconductivity}
		\begin{figure}
			\centering
			\includegraphics[width=0.5\linewidth]{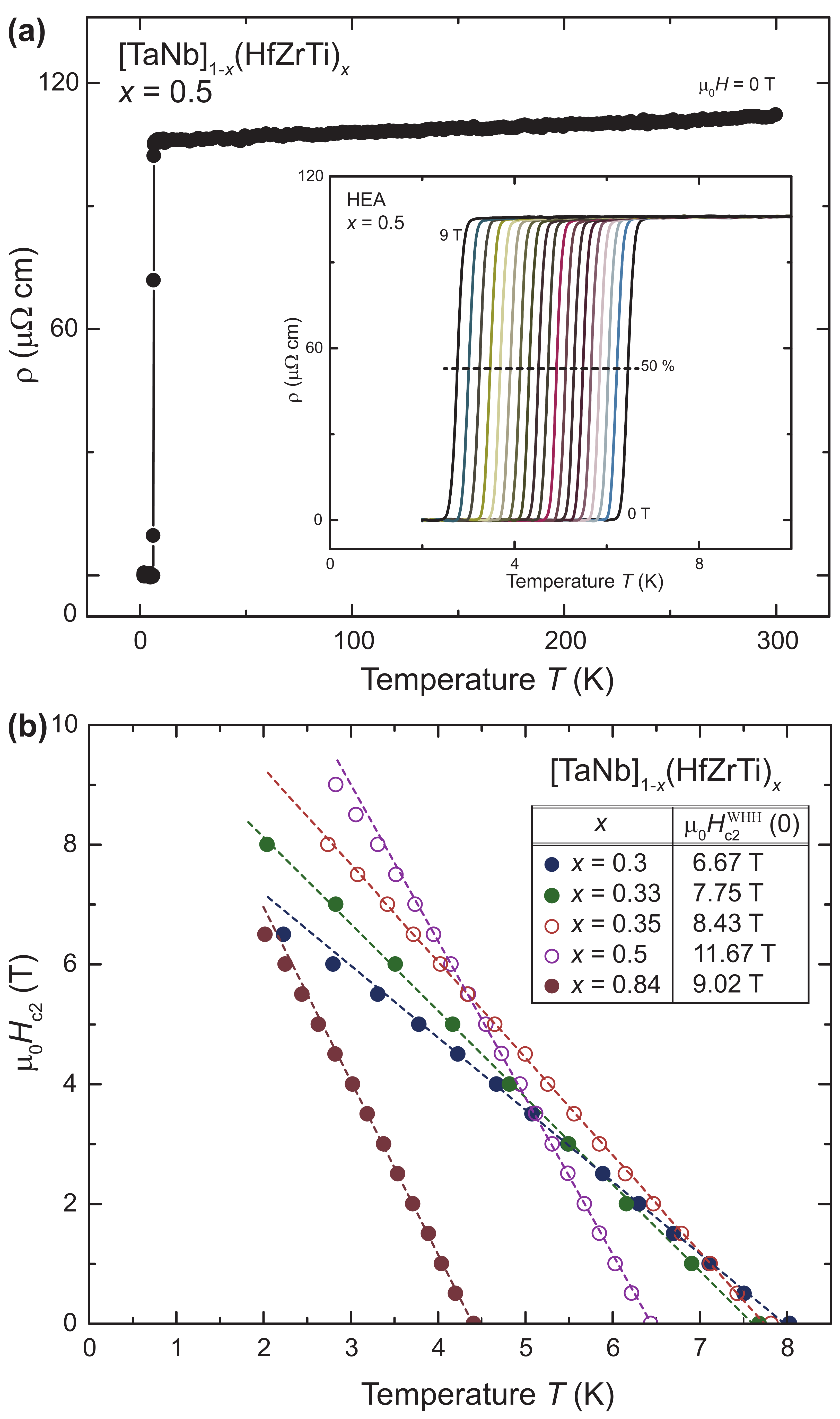}
			\caption{(a) Resistivity between 2 and 300 K of the HEA \ce{[TaNb]_{1-\textit{x}}(ZrHfTi)_{\textit{x}}} with \textit{x} = 0.5. In the inset the magnetic field dependence of the superconducting transition is shown in fields between $\mu_0H =$ 0 T to 9 T in steps of 0.5 T. The dotted line displays the 50 \% criterion, which is commonly used for the determination of the upper critical field $H_{\rm c2}$. (b) Temperature dependence of the upper critical field $H_{\rm c2}$, determined by the 50 \% criterion, of the  \ce{[TaNb]_{1-\textit{x}}(ZrHfTi)_{\textit{x}}} HEAs with \textit{x} = 0.30, 0.33, 0.35, 0.5, and 0.84. The lines are linear fits for determination of $(dH_{\rm c2}/dT)_{T = T_{\rm c}}$.}
			\label{fig:3}
		\end{figure}
%
			\begin{table}
				\begin{center}
					\begin{tabular}{| c | c | c | c |}
						\hline
						\ \ \ce{[TaNb]_{1-\textit{x}}(ZrHfTi)_{\textit{x}}} \ \ & \ \ $T_{\rm c}$ (resistivity) [K] \ \ & \ \ $\left(\frac{dH_{\rm c2}}{dT}\right)_{T = T_{\rm c}}$ [T/K] \ \ & \ \ $\mu_0 H^{\rm WHH}_{\rm c2}$(0) [T] \ \ \\
						\hline \hline
						$x$ = 0.3 & 8.03  & -1.203 & 6.67 \\
						$x$ = 0.33 & 7.75  & -1.449 & 7.75 \\
						$x$ = 0.4 & 7.56  & -1.616 & 8.43 \\
						$x$ = 0.5 & 6.46  & -2.618 & 11.67 \\
						$x$ = 0.84 & 4.52  & -2.893 & 9.02 \\
						\hline
					\end{tabular}
					\caption{Critical temperatures $T_{\rm c}$, slopes of the upper critical fields $\left(\frac{dH_{\rm c2}}{dT}\right)_{T = T_{\rm c}}$, and upper critical fields at zero temperature $\mu_0 H^{\rm WHH}_{\rm c2}$(0) of \ce{[TaNb]_{1-\textit{x}}(ZrHfTi)_{\textit{x}}} with $x$ = 0.3, 0.33, 0.4, 0.5, and 0.84}
					\label{tab:1}
				\end{center}
			\end{table}
	In figure \ref{fig:2a}, we show the zero-field cooled (ZFC) magnetization of the HEAs \ce{[TaNb]_{1-\textit{x}}(ZrHfTi)_{\textit{x}}} with \textit{x} = 0.2, 0.25, 0.3, 0.33, 0.35, 0.4, 0.45, 0.5, 0.6, 0.7, 0.8, and 0.84. The measurements were performed between 1.8 to 9 K, with zero-field cooling and in an external field of $\mu_0 H =$ 2 mT. For all samples a susceptibility larger than $\chi \approx -1$ ($-$1 is the ideal value for a fully superconducting material) below $T_{\rm c}$ was observed (the values more negative than $\chi = -1$  are caused by demagnetization effects). The temperature-dependent magnetizations are therefore plotted as $M(T)$/$M$(0) for better comparability. The superconducting phase transitions of all samples are well defined in temperature. The critical temperatures $T_{\rm c}$ were determined as the values at the points where the linearly approximated slopes (dashed line) cross the normal state magnetization, as illustrated by arrows in figure \ref{fig:2a} for example for the sample \textit{x} = 0.6.\footnote{No change in the general trend of critical temperatures is observed when the inflection point of the transition in $M$($T$) is used to define $T_{\rm c}$.} The critical temperatures $T_{\rm c}$  of the HEAs are plotted in figure \ref{fig:2b} as a function of the electron/atom (e/a) ratio (purple squares, blue line is a trend line; for a review on electron counting, see reference \onlinecite{Claudia}). For comparison the observed trend lines of the critical temperatures of the transition metals and their alloys in the crystalline form \cite{Matthias} (yellow dashed line) and as amorphous vapor-deposited films \cite{amorph1} (red dashed line) are also depicted. The trend of transition metals is often referred to as the Matthias rule, which links the $T_{\rm c}$ maxima with the non-integer d-electron count in simple binary alloys \cite{Matthias,Simon}. The trend-line for amorphous superconductors is from the pioneering work of Collver and Hammond \cite{amorph1, amorph2, Book_alloys}, who studied the critical temperature $T_{\rm c}$ of vapor-cryodeposited films of transition-metal alloys and came to the conclusion that $T_{\rm c}$ versus electron/atom ratio no longer exhibited the characteristic behavior of the Matthias rule for crystalline binary alloys. Instead they found that the critical temperatures $T_{\rm c}$ increase with increasing e/a, in a monotonic, rather structureless way with a maximum at a much higher e/a(d-electrons) = 6.4. These two curves, the Matthias rule, and the amorphous critical temperatures $T_{\rm c}$ after Collver and Hammond are the established standards to which other superconductors may be compared. Both of these trend lines have been the subject of extensive theoretical research as well \cite{Book_alloys}. \\ \\
	The critical temperatures $T_{\rm c}$ of the HEA \ce{[TaNb]_{1-\textit{x}}(ZrHfTi)_{\textit{x}}} fall in between the two benchmark lines. The increase of the transition temperatures is clearly less pronounced than for crystalline alloys, and follows rather a monotonically increasing trend as is observed for the amorphous superconductors. However, a maximum is reached near e/a(d-electrons) = 4.7, which is an essential feature of the Matthias rule, even though the maximum is much broader for the simple crystalline superconductors. Therefore, the great disorder of the HEA gives us the opportunity to investigate a superconducting system between the crystalline and amorphous benchmarks, with distinct features of both. Even though all chemical compositions used for this study are within a broad definition of HEAs (see above), the mixing entropy $\Delta S_{\rm mixing}$ changes nevertheless across the series. The mixing entropy $\Delta S_{\rm mixing}$ is estimated as is commonly done for HEAs assuming a mixture of hard spheres, in accordance with Mansoori \textit{et al}. \cite{Mansoori,HEA_review} The largest mixing entropy $\Delta S_{\rm mixing}$ is present at a ratio of e/a(d-electrons) = 4.4. The HEA series \ce{[TaNb]_{1-\textit{x}}(ZrHfTi)_{\textit{x}}} can therefore additionally be interpreted as a solid solution ranging from a higher mixing entropy to a lower one. This may explain the general trends across the series: In the region of a more amorphous-like increase of the transition temperatures $T_{\rm c}$ the highest mixing entropy is present, while in the region of the phase diagram with the lowest mixing entropy a maximum of the transition temperatures $T_{\rm c}$, which is in agreement with the Matthias rule, is observed.
	\subsection*{Upper Critical Fields \ce{\textit{H}_{c2}} of the HEAs and the Effect of Increasing Mixing Entropy}
	In figure \ref{fig:3}a, we show the temperature dependent electrical resistivity $\rho$ of the HEA \ce{[TaNb]_{1-\textit{x}}(ZrHfTi)_{\textit{x}}} with \textit{x} = 0.5 in a temperature range between $T =$ 2 and 300 K. The resistivity at room temperature exhibits a value of $\rho({\rm 300 \ K}) \approx$ 116 $\mu\Omega \rm{cm}$. The resistivity is found to be metallic and decreasing linearly with decreasing temperature. The residual resistivity ratio (RRR), ${\rm RRR} = rho(300 {\rm K})/rho(8 {\rm K})) \approx 1.1$ is a low value, comparable to that observed for nonstoichiometric or highly disordered intermetallic compounds. The linear behavior of $\rho(T)$ is also a common behavior for highly disordered alloys, caused by the short lifetimes of the quasiparticles, which are scattered by the disorder and therefore decohere. This kind of conductivity is generally referred to as "bad metal conductivity". It is also found in strongly correlated materials such as the high $T_{\rm c}$-superconductors, and in transition metal systems, e.g. \ce{VO2} \cite{badmetal1,badmetal2}. In the inset of figure \ref{fig:3}a the magnetic field dependence of the resistivity in the vicinity of the superconducting phase transition is shown for the sample with \textit{x} = 0.5, for magnetic fields between $\mu_0H =$ 0 T to 9 T in 0.5 T steps. The transition temperature $T_{\rm c}$ is reduced with increasing field $H$. The superconductivity is at 9 T still observable above $T$ = 2 K, indicating a high upper critical field $\mu_0H_{\rm c2}(0)$. (The upper critical fields were determined by the 50 \% criterion, i.e., the upper critical field $\mu_0H_{c2}(T)$ is defined by the temperature $T$ at which 50 \% of the normal-state resistivity is suppressed, as illustrated by the dashed line in figures \ref{fig:3}a (see, e.g., references \onlinecite{Hc2,Hc2_2,Hc2_3})). In figure \ref{fig:3}b, the temperature dependence of the upper critical fields $\mu_0H_{\rm c2}(T)$ of the HEAs \ce{[TaNb]_{1-\textit{x}}(ZrHfTi)_{\textit{x}}} with \textit{x} = 0.30, 0.33, 0.35, 0.5, and 0.84 are shown. The dashed lines in figure \ref{fig:3}b represent the slopes of the upper critical fields $\left(\frac{dH_{\rm c2}}{dT}\right)_{T = T_{\rm c}}$ for all five samples, respectively. These slopes are used to estimate the upper critical fields at zero temperature $\mu_0 H_{c2}$(0) by applying the Werthamer-Helfand-Hohenberg (WHH) approximation in the dirty limit \cite{WHH}, according to,
	\begin{equation}
	H^{\rm WHH}_{c2}(0) = -0.69 \ T_{\rm c} \ \left(\frac{dH_{\rm c2}}{dT}\right)_{T = T_c}.
	\label{eq:WHH}
	\end{equation}
	The obtained critical temperatures (from the resistivity measurements), the slopes of the upper critical fields $\left(\frac{dH_{\rm c2}}{dT}\right)_{T = T_{\rm c}}$, and estimated values after WHH of the upper critical fields at zero temperature $\mu_0 H^{\rm WHH}_{\rm c2}$(0) are summarized in table \ref{tab:1}. It is noteworthy that the slopes of the upper critical field increase with increasing mixing entropy of the system. Therefore, it is not the member of this series with the highest critical temperature $T_{\rm c}$ that has the largest $\mu_0 H_{\rm c2}$(0). Rather, the sample \textit{x} = 0.5 has the largest upper critical field with a large negative slope of the upper critical field $\left(\frac{dH_{c2}}{dT}\right)_{T = T_c} \approx$ -2.618 T/K and an overall upper critical field $\mu_0 H_{\rm c2}$(0) $\approx$ 11.67 T. This value is very close to the the Pauli paramagnetic limit $\mu_0 H^{\rm Pauli}_{\rm c2} = 1.84 \ T_{\rm c} = 11.9$ T. For the sample with $x$ = 0.84, the slope of the upper critical field is $\left(\frac{dH_{\rm c2}}{dT}\right)_{T = T_c} \approx$ -2.893 T/K, the largest in absolute value. This sample is also the one with the largest mixing entropy $\Delta S_{\rm mixing}$ among the investigated alloys. For $x$ = 0.84, the experimentally observed upper critical field $\mu_0 H_{\rm c2}$(0) is even found to exceed the Pauli paramagnetic limit $\mu_0 H^{\rm Pauli}_{\rm c2} = 1.84 \ T_{\rm c} = 8.3$ T. Therefore, strong spin-orbit coupling may play a role in the characteristic properties of the superconducting state in these HEAs. The observed systematic change of $\mu_0 H_{\rm c2}$(0) does not, however, correlate with the atomic spin-orbit coupling, which does not change much in the series, and therefore a relationship between $\mu_0 H_{\rm c2}$(0)and the magnitude of spin orbit coupling cannot be established here. \\
			\begin{figure}
				\centering
				\includegraphics[width=0.8\linewidth]{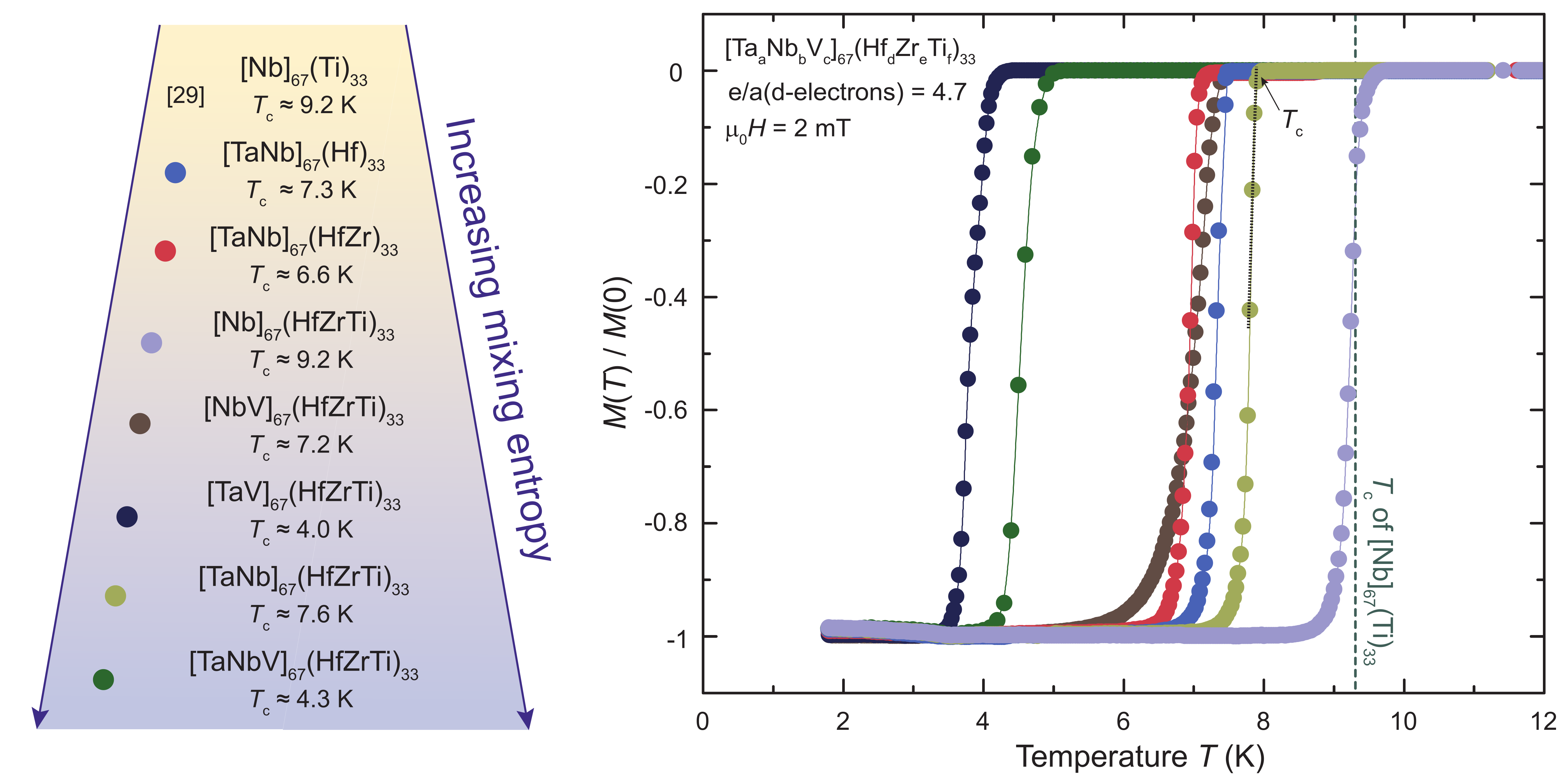}
				\caption{The effect of alloy complexity on the superconducting transition. The ZFC magnetization in an external field of $\mu_0 H =$ 2 mT of the alloys \ce{[TaNb]_{0.67}(ZrHfTi)_{0.33}}, \ce{[Nb]_{0.67}(ZrHfTi)_{0.33}}, \ce{[TaNb]_{0.67}(ZrHf)_{0.33}}, \ce{[TaNb]_{0.67}(Hf)_{0.33}}, and \ce{[TaNbV]_{0.67}(ZrHfTi)_{0.33}} in the vicinity of the superconducting transition.}
				\label{fig:4}
			\end{figure}
	The mixing entropy $\Delta S_{\rm mixing}$ can either be reduced by the method described above, or by the reduction of the number of constituent atoms of the alloy. We have prepared seven alloys for comparison, close to the optimal valence electron concentration of e/a(d-electron) = 4.7. These alloys are all found to randomly arrange on BCC lattices, as expected (see, e.g., references \onlinecite{Koc77,Book_alloys}). In figure \ref{fig:4}, we show the ZFC magnetization of the alloys \ce{[TaNbV]_{0.67}(ZrHfTi)_{0.33}}, \ce{[TaNb]_{0.67}(ZrHfTi)_{0.33}}, \ce{[TaV]_{0.67}(ZrHfTi)_{0.33}}, \ce{[NbV]_{0.67}(ZrHfTi)_{0.33}} \ce{[Nb]_{0.0.67}(ZrHfTi)_{0.33}}, \ce{[TaNb]_{0.67}(ZrHf)_{0.33}}, and \ce{[TaNb]_{0.67}(Hf)_{0.33}} in the vicinity of the superconducting transition measured in an external field of $\mu_0 H =$ 2 mT. The critical temperature $T_{\rm c}$ is found to decrease very little on going from the binary alloy \ce{[Nb]_{0.67}(Ti)_{0.33}}, with a critical temperature of $T_{\rm c} \approx$ 9.2 K \cite{Matthias,Koc77}, to the HEA \ce{[TaNb]_{0.67}(ZrHfTi)_{0.33}}, where the atoms are highly disordered. The disorder introduced by the increasing number of constituent atoms does not lead to a loss of the superconductivity or to a very large decrease of the critical temperatures $T_{\rm c}$. It is also apparent that the superconducting properties of these alloys are not just a compositional mixture of all the properties of the constitute elements, but rather that a single homogeneous superconducting phase is observed for all of them; the highly disordered atomic content of the alloy conspires to give rise to one homogeneous superconducting state. In this sense superconductivity in HEA is a logical further development of transition-metal alloys consisting of constituent atoms with a VEC of 4 and 5. The critical temperature decreases to $T_{\rm c} \approx$ 4.2 K for \ce{[TaNbV]_{0.67}(ZrHfTi)_{0.33}} is indicating that the elemental make-up is significant for the physical properties even for the highly disordered atoms on simple lattices in HEAs.
	\subsection*{Electron-Phonon Coupling in the HEA Superconductor}
	\begin{figure}
		\centering
		\includegraphics[width=0.48\linewidth]{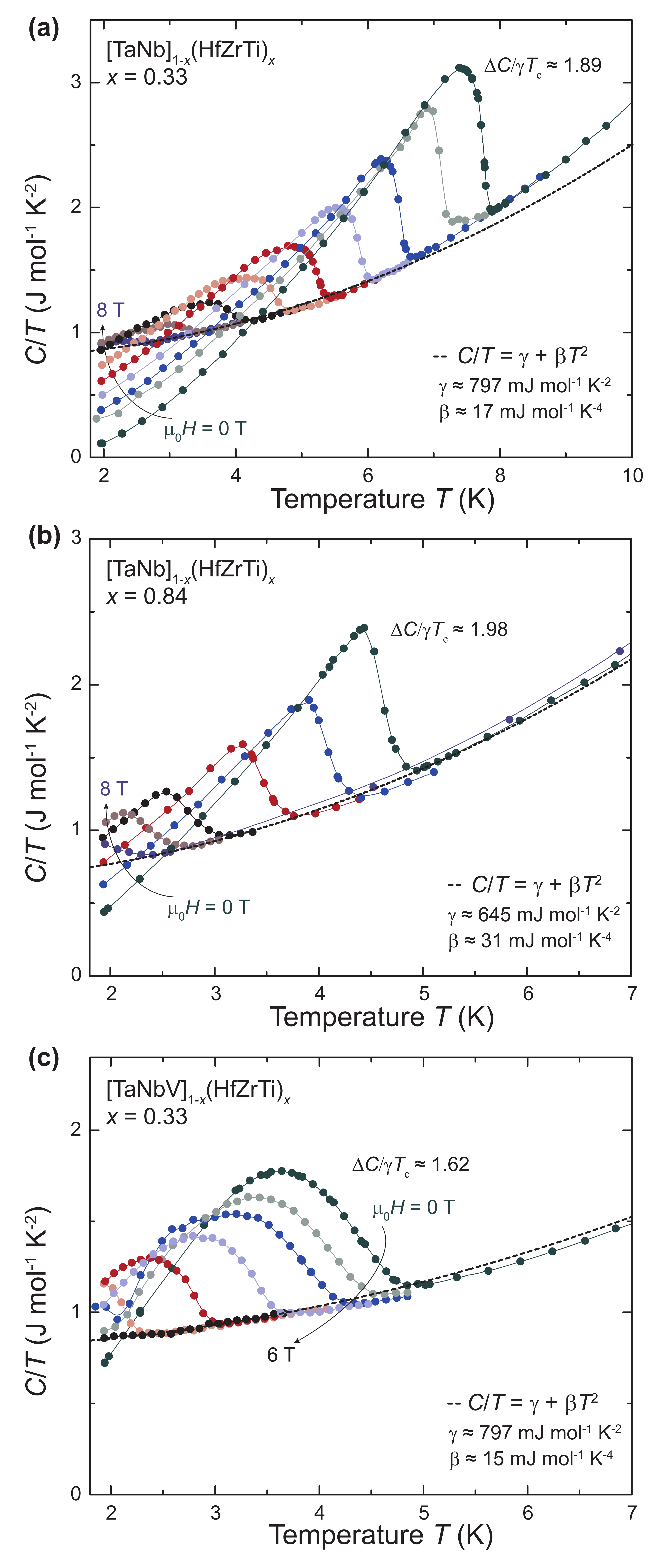}
		\caption{Specific heat measurements in fields from $\mu_0 H =$ 0 T to 8 T in the vicinity of the superconducting phase transition, for three representatives samples: (a) the nearly optimally doped HEA \ce{[TaNb]_{1-\textit{x}}(ZrHfTi)_{\textit{x}}} with \textit{x} = 0.33 (b) the HEA with \textit{x} = 0.84, which has a upper critical field $\mu_0 H_{\rm c2}$ above the Pauli limit, and (c) the nearly  optimally doped HEA \ce{[TaNbV]_{1-\textit{x}}(ZrHfTi)_{\textit{x}}} which includes vanadium, with \textit{x} = 0.33; the $T_{\rm c}$ for this HEA is significantly lower than for the equivalent electron-count HEA where vanadium is not present.}
		\label{fig:5}
	\end{figure}
	We have performed specific heat measurements on \ce{[TaNb]_{0.67}(ZrHfTi)_{0.33}}, \ce{[TaNb]_{0.16}(ZrHfTi)_{0.84}}, and \ce{[TaNbV]_{0.67}(ZrHfTi)_{0.33}}, in order to get further insights into the nature of the different critical temperatures $T_{\rm c}$ that are the result of varying electron count and elemental composition of the alloys. In figure \ref{fig:5}, we show the temperature-dependent specific heat capacities in fields from $\mu_0 H =$ 0 T to 8 T in the vicinity of the superconducting phase transition of the three alloys. All three alloys display a single well defined transition, which is further evidence for the emergence of a single collective superconducting phase. The normal-state contribution has been fitted to the data at low temperatures (dotted lines) according to  
	\begin{equation}
	\frac{C(T)}{T} = \gamma + \beta T^2
	\end{equation}  
					\begin{table}
						\begin{center}
							\begin{tabular}{| c || c | c | c |}
								\hline
								Specific Heat Par. & \ce{[TaNb]_{0.67}(ZrHfTi)_{0.33}} & \ce{[TaNb]_{0.16}(ZrHfTi)_{0.84}}  & \ce{[TaNbV]_{0.67}(ZrHfTi)_{0.33}} \\
								\hline \hline
								\ \ $T_{\rm c}$ [K] \ \ 					& 7.70  & 4.59 & 4.11 \\
								\ \ $\gamma$ [mJ mol$^{-1}$ K$^{-2}$] \ \ 	& 7.97(5)  & 6.45(8) & 7.97(4) \\
								\ \ $\beta$ [mJ mol$^{-1}$ K$^{-4}$] \ \ 	& 0.170(4)  & 0.311(9) & 1.48(5) \\
								\ \ $\Theta_{\rm D}$ [K] \ \ 				& 225(2)  & 184(2) & 236(3) \\
								\ \ $\lambda_{\rm el-ph}$ \ \ 				& 0.83    & 0.73  & 0.65 \\
								$D(E_{\rm F})$ [st. eV$^{-1}$/at. f.u.] 	& 1.9    & 1.6  & 2.1 \\
								\ \ $\Delta C / \gamma T_{\rm c}$ \ \ 		& 1.89  & 1.98 & 1.62 \\
								\ \ $\Delta$(0) [meV]                       &	1.21(2) &	0.71(1) &	0.56(1) \\
								\ \ 2$\Delta$(0)/k$_{\rm B} T_{\rm c}$      & 3.7 &	3.6 & 	3.2 \\
								
								\hline
							\end{tabular}
							\caption{Summary of the electronic and phononic contributions to the superconductivity in the HEAs. The error of the fit of the specific heat is given in brackets.}
							\label{tab:2}
						\end{center}
					\end{table}
	with the Sommerfeld constant $\gamma$ and $\beta = 12 \pi^4 n R / 5 \Theta_{\rm D}^3$, where $n$ is the number of atoms per formula unit, $R$ = 8.31 J mol$^{-1}$ K$^{-1}$ is the gas constant, and $\Theta_D$ the Debye temperature. For comparability reasons the numbers of atoms per formula unit $n$ was fixed to be one hundred. The obtained values for $\gamma$, $\beta$, and $\Theta_D$ are summarized in table \ref{tab:2}. The obtained ratios $\Delta C / \gamma T_{\rm c}$ all exceed the standard weak-coupling BCS value, which is $\Delta C / \gamma T_c =$ 1.43, indicating intermediate to strong-coupling superconductivity. The Sommerfeld constant is found to decrease substantially from $\gamma \approx$ 797 mJ mol$^{-1}$ K$^{-2}$ to 645 mJ mol$^{-1}$ K$^{-2}$ with a decreasing electron count within the series \ce{[TaNb]_{1-\textit{x}}(ZrHfTi)_{\textit{x}}}. Thereby, the density of states at the Fermi-level is reduced, since $\gamma$ is proportional to the density of states at the Fermi-level ($\gamma \propto D(E_{\rm F})$). Thus we attribute the decreasing of the critical temperature $T_{\rm c}$ with an increasing electron count to a significant decrease in the density of states. It should be noted that simultaneously also the electron-phonon coupling is lowered, which also might contribute to the lowering of the critical temperature (see below). \ce{[TaNb]_{0.67}(ZrHfTi)_{0.33}} and \ce{[TaNbV]_{0.67}(ZrHfTi)_{0.33}} have nominally the same electron count, and experimentally we find the same density of states at the Fermi-level, with almost identical values for $\gamma$. The Debye temperature is found to increase only slightly. Thus although the critical temperatures differ by almost a factor of 2, there is not much difference in the fundamental quantities that determine the transition temperature: $\gamma$ and $\Theta_{\rm D}$. We therefore tentatively attribute the decrease in $T_{\rm c}$ to the difference in the electron-phonon coupling $\lambda$ that must occur on going from \ce{[TaNb]_{0.67}(ZrHfTi)_{0.33}} to \ce{[TaNbV]_{0.67}(ZrHfTi)_{0.33}}. The electron-phonon coupling $\lambda_{\rm el-ph}$ can be estimated using the approximated McMillan formula, which is based on the phonon spectrum of niobium \cite{McMillan,Dynes72} and is valid for $\lambda < 1.25$ \cite{Dynes75}:
	\begin{equation}
	\lambda_{\rm el-ph} = \dfrac{1.04 + \mu^{*} \ {\rm ln}\big(\frac{\Theta_{\rm D}}{1.45 T_{\rm c}}\big)}{(1-0.62 \mu^{*}) {\rm ln}\big(\frac{\Theta_{\rm D}}{1.45 T_{\rm c}}\big)-1.04}
	\end{equation} 
	The parameter $\mu^{*}$ is the effective Coulomb repulsion which arises from Coulomb-coupling propagating much more rapidly than phonon-coupling. Here, we are using a value of $\mu^{*}$ = 0.13, which is an average value used commonly for intermetallic superconductors (see, e.g., reference \onlinecite{Tomasz}). Having the Sommerfeld parameter and the electron-phonon coupling, the noninteracting density of states at the Fermi energy can be calculated according to:
	\begin{equation}
	D(E_{\rm F}) = \dfrac{3 \gamma}{\pi^2 k_{\rm B}^2 (1+\lambda_{\rm el-ph})}.
	\end{equation}
	From the electronic low temperature specific heat data, we have estimated the value of the superconducting gap of all three compounds, according to
	\begin{equation}
	C_{\rm el} = a \ exp(-\Delta(0)/k_{\rm B}T_c).
	\end{equation}
	The obtained values for the electronic and phononic contributions to the superconductivity in HEAs are summarized in table \ref{tab:2}. The values for $\Delta$(0) are similar to comparable intermetallic superconductors and for all three samples the value for 2$\Delta$(0)/k$_{\rm B} T_{\rm c}$ is close to the expected value of 3.52, which is expected for s-wave superconductors according to the BCS model. The estimated values for the electron-phonon coupling $\lambda_{\rm el-ph}$ further support that the density of states at the Fermi level $D(E_{\rm F})$ remains the same for \ce{[TaNb]_{0.67}(ZrHfTi)_{0.33}} and \ce{[TaNbV]_{0.67}(ZrHfTi)_{0.33}}, while the electron-phonon coupling constant $\lambda_{\rm el-ph}$ is strongly reduced for the latter material. This finding supports the general concept that specific elements are essential for optimized superconductivity in compounds. Here we find that the elemental make-up is crucial even in the case of a highly disordered multi-compontent HEA superconductor.
	\section*{Summary and conclusion}
	We have synthesized the HEA \ce{[TaNb]_{1-\textit{x}}(ZrHfTi)_{\textit{x}}} for \textit{x} = 0.2, 0.25, 0.3, 0.33, 0.35, 0.4, 0.45, 0.5, 0.6, 0.7, 0.8, and 0.84 by arcmelting of the elements under argon and by subsequent quenching. We found from x-ray powder diffraction measurements that all these alloys arrange on a simple BCC crystal lattice ($Im\bar{3}m$), with unit cell parameters between $a_0 \approx 3.334$ \AA \ to $3.431$~\AA \ within the solid solution. All prepared samples are found to be bulk superconductors with critical temperatures between $T_{\rm c} \approx$ 4.49 K and 7.92 K. By comparison of the critical temperatures of \ce{[TaNb]_{1-\textit{x}}(ZrHfTi)_{\textit{x}}} with the critical temperatures of the transition metals and their alloys in the crystalline form and as amorphous vapor-deposited films, we find the superconducting HEA to display characteristics intermediate to both of them. The valence electron dependence of the transition temperatures for \ce{[TaNb]_{1-\textit{x}}(ZrHfTi)_{\textit{x}}} is clearly less pronounced than that seen for crystalline alloys. However, a maximum is reached right around e/a(d-electron) = 4.7, which is an essential feature of the Matthias rule for crystalline transition metal superconductors. Therefore, we find that this system neither follows a crystalline nor an amorphous-like trend for this collective electron state. We find the temperature dependent electrical resistivity $\rho$ of the HEAs \ce{[TaNb]_{1-\textit{x}}(ZrHfTi)_{\textit{x}}} to be metallic and decreasing linearly with decreasing temperature and that the slopes of the upper critical field $\left(\frac{dH_{\rm c2}}{dT}\right)_{T = T_{\rm c}}$ increase with increasing mixing entropy of the system. It is, therefore, not the member of this series with the highest critical temperature $T_{\rm c}$ that has the largest $\mu_0 H_{c2}$(0). Rather, the sample with \textit{x} = 0.5 has the largest upper critical field, with a large negative slope of the upper critical field $\left(\frac{dH_{c2}}{dT}\right)_{T = T_c} \approx$ -2.618 T/K and an overall $\mu_0 H_{c2}$(0) $\approx$ 11.67 T. By reducing the mixing entropy $\Delta S_{\rm mixing}$ the critical temperatures are found to decrease only slightly from the binary alloy \ce{[Nb]_{0.67}(Ti)_{0.33}} with a critical temperature of $T_{\rm c} \approx$ 9.2 K to the HEA \ce{[TaNb]_{0.67}(ZrHfTi)_{0.33}}. Thus the disorder introduced by the increasing number of constituent atoms does not lead to a loss of the superconductivity or a large decrease of the critical temperature $T_{\rm c}$. We do find however that the effect of elemental make-up is significant for the physical properties even for the highly disordered atoms on the simple lattice in this superconducting HEA. The general interplay of chemical structure, disorder, and superconductivity are topics of fundamental interest. Many known superconductors are posed near structural instabilities, for example, the bismuth oxide superconductors \cite{BiO_SC,BiO_SC2}, the tungsten bronzes \cite{WO3}, and also many intermetallic superconductors \cite{Daigo,A15}. The superconducting HEA studied here offers the unique opportunity to investigate superconductivity on one of the three most fundamental crystal lattices stabilized by high-entropy mixing. For future work, it will be of interest to determine the electronic and phononic densities of states of these alloys in order to understand their interplay. Our results suggest that HEAs are versatile model systems for the investigation of structure-property relations, as well as for the understanding of the change of electronic properties, on going from crystalline to amorphous superconducting materials.
	\section*{Acknowledgments}
	This work was primarily supported by the Gordon and Betty Moore Foundation, EPiQS initiative, Grant GBMF-4412. The research performed at the Gdansk University of Technology was financially supported by the National Science Centre (Poland) Grant No. DEC-2012/07/E/ST3/00584. The electron microscope work done at Brookhaven National Laboratory was supported by the DOE BES, by the Materials Sciences and Engineering Division, under Contract DE-AC02-98CH10886.
\end{document}